\documentclass[fleqn,10pt]{wlscirep}
\title{Ionic transport through a protein nanopore: a Coarse-Grained Molecular Dynamics Study}

\author[1]{Nathalie Basdevant}
\author[1]{Delphine Dessaux}
\author[1,*]{Rosa Ramirez}
\affil[1]{LAMBE, Univ Evry, CNRS, CEA, Universit\'e Paris-Saclay, 91025, Evry, France }

\affil[*]{rosa.ramirez@univ-evry.fr}

\usepackage{placeins}

\begin{abstract}
The MARTINI coarse-grained (CG) force field is used to test the ability of CG models to simulate ionic transport through protein nanopores. The ionic conductivity of CG ions in solution was computed and compared with experimental results. Next, we studied the electrostatic behavior of a solvated CG lipid bilayer in salt solution under an external electric field. We showed this approach correctly describes the experimental conditions under a potential bias. Finally, we performed CG molecular dynamics simulations of the ionic transport through a protein nanopore ($\alpha$-hemolysin) inserted in a lipid bilayer, under different electric fields, for 2-3 microseconds. The resulting $I-V$ curve is qualitatively consistent with experiments, although the computed current is one order of magnitude smaller. Current saturation was observed for potential biases over $\pm~350$~mV. We also discuss the time to reach a stationary regime and the role of the protein flexibility in our CG simulations. 

\end{abstract}
\begin{document}

\flushbottom
\maketitle

\thispagestyle{empty}

\section*{Introduction}
Single-molecule experimental techniques are widely used to understand the kinetics and energetics involved into biological phenomena. Among them, because of its simplicity and amenability to parallelization, nanopore force spectroscopy is an increasingly used technique for ultra-sensitive detection of macromolecules. The method consists in applying an electric potential difference to guide a charged molecule through a nanopore (artificial or biological) inserted in a solid or lipid membrane, in presence of a salt solution~\cite{Kasianowicz1996, Dudko2010, Meller2000, Oukhaled2007, Oukhaled2008}. A macromolecule going through the pore, partially blocking it, creates a variation in the electrical current according to the nature of the molecule and the pore characteristics: size, conformation, and structure. Ionic current measurements are a probe for the detection. Amongst the various biological nanopore, $\alpha$-hemolysin is the most widely employed in nanopore analytics. It has been commonly used as a nanopore for DNA or RNA translocation~\cite{Kasianowicz1996,Meller2000}, DNA sequencing or biosensing~\cite{Deamer2016}, DNA unzipping of individual hairpins~\cite{Mathe2004,Mathe2005,Muzard2010}, and the study of protein translocation~\cite{Oukhaled2007}. 

The understanding of the physical mechanisms governing the translocation of ions or molecules through a nanopore is not only helpful to reveal relevant biological phenomena but also for the design of new devices based on such nanopores. The microscopic details of these processes cannot be deduced straightforwardly from the experiments. In this sense, an increasing effort to provide theoretical and computational approaches in order to understand such processes is being done~\cite{Modi2012}. In general, these approaches require a choice between the amount of computational resources and the microscopic level of accuracy. Among continuum approaches, the Poisson–Nernst–Planck theory (PNP)~\cite{Roux2004, Noskov2004, Cervera2005, Constantin2007, Cozmuta2004, Birlea2012} is useful to study ion transport at a relatively small computational cost, although not very accurate for narrow pores. At the same level, Brownian or Langevin dynamics methods (BD)~\cite{DeBiase2014, DeBiase2016, Millar2008, Noskov2004} can be an alternative. In this case, the main challenge is the representation of the protein-continuum solvent interface.
 
In All-Atom Molecular Dynamics (AA-MD) simulations, the individual motion of ions, water, protein and membrane atoms are fully computed based on a classical atomic force field. This approach will be limited by the classical force field approximation used, but AA-MD simulations are undoubtedly excellent to provide microscopic details of biological translocation processes "at short times". The huge computational cost is the main disadvantage of this method and therefore, there are several ways to accelerate the transport of molecules through the $\alpha$-hemolysin pore: using very high electric potentials (1~V compared to hundreds of mV for experiments)~\cite{Aksimentiev2004, Aksimentiev2005, Mathe2005} or forcing the transport of molecules through the pore using velocity-pulling methods~\cite{DiMarino2015}. Although some results of these AA-MD are in accordance with experiments, long time phenomena cannot be observed in such simulations, generally limited to a few hundred nanoseconds. The high-potential/pulling methods can also induce artifacts in the results.

Coarse-Grained (CG) force fields, representing several atoms by one site, for lipid membrane and biomolecules, are a very promising alternative to All-Atoms models, as they enable simulations of time-lengths closer to experiments. In this sense, our article focuses on MD simulations where the AA representation is replaced by an "atomistic" CG force field, in our case the MARTINI force field. The CG MARTINI force field~\cite{Marrink2007}, originally developed for lipids, is widely used to simulate membrane systems for time-lengths up to hundreds of microseconds~\cite{Ingolfsson2014,Pinot2014,VanEerden2015}. Furthermore, a polarizable CG water model is needed to correctly describe the electrostatic behavior of electrolytes.  Among the several CG polarizable solvent models available~\cite{Ha-Duong2009,Wu2010,Yesylevskyy2010}, the Polarizable Water (PW) model~\cite{Yesylevskyy2010} has been parametrized to be used together with the standard MARTINI force field. Whereas classical AA simulations of ionic current through $\alpha$-hemolysin have been done, to our knowledge, such a study using a CG model has not been yet performed. 

In this work, our MARTINI CG MD simulations for determining ion currents through $\alpha$-hemolysin are detailed and discussed. We begin by studying the electrostatic behavior of solvent, ions and membrane using this force field. The ionic conductivity of CG MARTINI ions in PW solution as a function of concentration is discussed. Secondly, we study the electrostatic behavior of a solvated DPPC (dipalmitoylphosphatidylcholine) lipid bilayer with excess ions in solution. We will confirm that applying an electric field in a direction perpendicular to the membrane plan is accurate to represent the experimentally applied potential difference. Finally, we will study the current through the protein nanopore at various electric potentials on microseconds-long MD.

\section*{Results}

All our MD simulations were performed using the Gromacs software package~\cite{VanDerSpoel2005} and the MARTINI 2.2 Coarse-Grained (CG) model for ions, lipids~\cite{Marrink2007}, proteins~\cite{Monticelli2008,Jong2013} and the PW (Polarizable Water) model~\cite{Yesylevskyy2010} for water. The systems were simulated either in the NVT or the NPT ensemble at a temperature of $T=320$~K, where PW water presents a density of 1020 kg/$m^3$~\cite{Yesylevskyy2010}. MARTINI CG ions consist in particles of the same size and mass as a PW particle, carrying positive or negative charge. Although they are called "Na$^+$" and "Cl$^-$" in the force field parameters files, they do not represent a realistic sodium or chlorine ion \cite{Marrink2004}. Since the MARTINI salt is perfectly symmetrical, we considered that it represents the KCl salt instead of NaCl. Methodological details of the simulations can be found in the Methods.The details of the ionic molar concentration setting for NVT simulations is discussed in the Supplementary information. 

\subsection*{Ionic conductivity}

In first place, we focused on the ionic conductivity of MARTINI ions in bulk PW water. The correct interplay between ions and water force fields is of critical importance to simulate conduction processes. In the case of MARTINI PW water, bulk properties such as density and dielectric constant have been optimized to construct the PW force field, while for ions the equilibrium bulk correlation function ION-PW water has been benchmarked to determine ions parameters~\cite{Yesylevskyy2010}. We would like to address the ability of the combination of these ion and water force fields to yield reasonable results when used to simulate transport processes. 

Our simulations consisted on a cubic box containing PW water particles, cations (K$^{+}$) and anions (Cl$^-$) with concentrations from 0.1~M to 2~M (see Methods for details). For each concentration, three different conditions were simulated: no external electric field, and electric fields of 0.02~V/nm and 0.04~V/nm in the $z$ direction.

\begin{figure}[htb]
\centering
\includegraphics[width=0.5\linewidth]{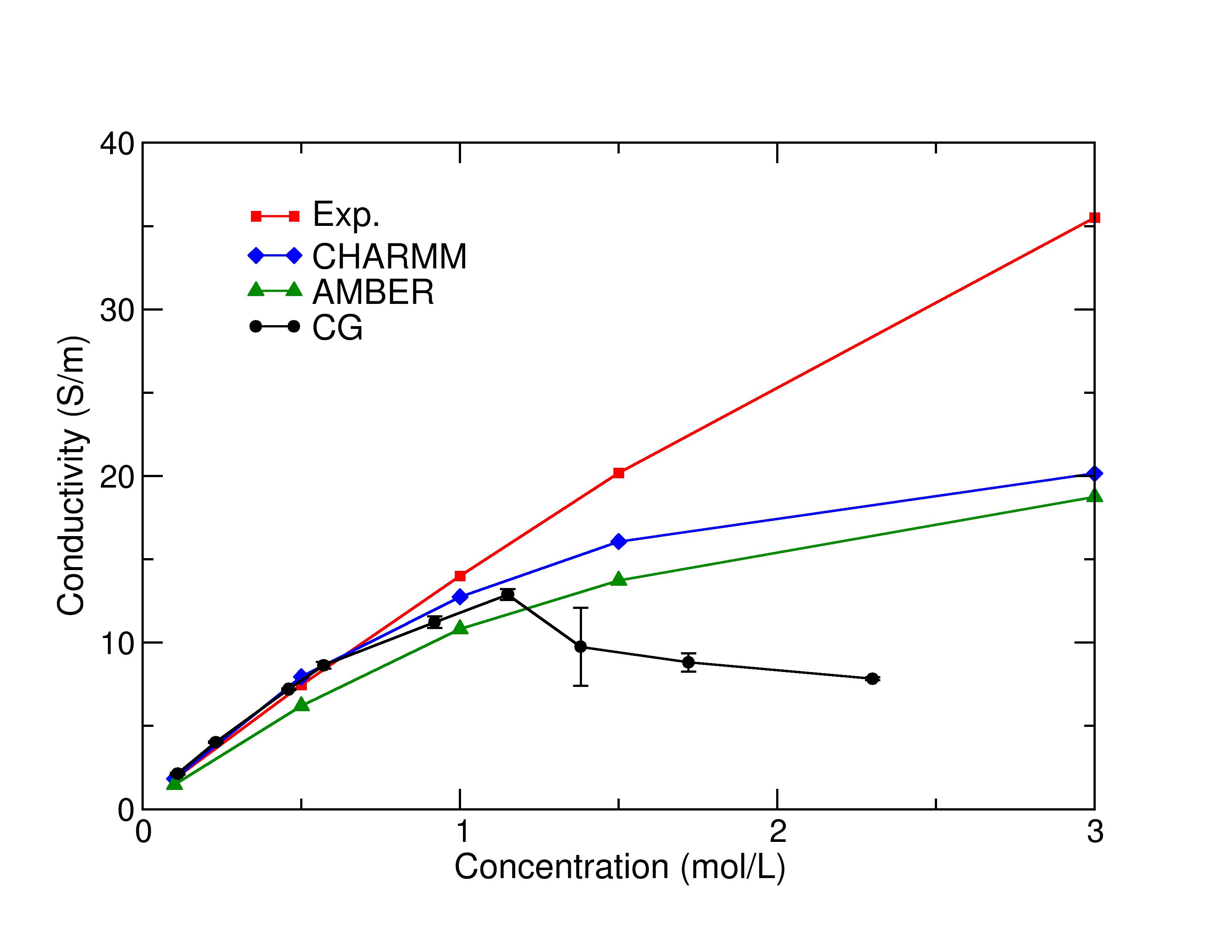}
\caption{Ionic conductivity of KCl as a function of molarity. Experimental results (in red) and all-atom simulations using the CHARMM (blue) and AMBER (green) force field performed by Pezeshki et al~\cite{Pezeshki2009} for KCl at 320~K, and CG results (black) computed with our CG MARTINI MD simulations.}
\label{fig:KClconduc}
\end{figure} 
To calculate the ionic conductivity $\kappa$ of CG ions as a function of concentration, we computed the average total ionic current $<I>$ in the $z$ direction as follows: the $z$-component of the velocity of the center of mass of all K$^{+}$ ions, $v^{+}_z$, and the one of all Cl$^-$ ions, $v^{-}_z$, were calculated every 500~ps, and the instantaneous current in the $z$ direction was then computed using the formula: 

\begin{eqnarray}
I(t) = \frac{1}{L_z} N_{ions} e (v^{+}_z-v^{-}_z) ,
\end{eqnarray}
where $N_{ions}$ is the total number of ions of each category, and $e$ is the elementary charge. The cumulative sum of the current over the length of the simulation (1~$\mu$s) was then calculated and fitted with a linear regression, the resulting slope of this fit yielding the average current $<I>$. The conductivity $\kappa$ is defined by $\kappa=\frac{<I>}{E A}$, where $A$ is the cross-sectional area, which is here: $A=L_xL_y$ ($L_x$ and $L_y$ are the boxes dimensions in $x$ and $y$ directions). Since two distinct electric fields were applied in the simulations at each concentration, we computed the average of the two calculated conductivities.

The concentration dependence of the conductivity computed at a fixed temperature of 320~K is shown on Fig~\ref{fig:KClconduc}, as well as the experimental results and AA simulations using the CHARMM and AMBER force field performed by Pezeshki \textit{et al}~\cite{Pezeshki2009,Lu2013} for KCl (numerical values from the figure of ref.~\cite{Pezeshki2009} were kindly provided by Ulrich Kleinekath\"ofer). The CG conductivity increases with the ionic concentration for concentrations below 1.15~M and is slightly lower than the reference values. For concentrations above this value, the conductivity decreases with the concentration. This effect can be explained by the size of the CG particles: at concentrations larger than 1.15~M, two opposite CG ions are separated by very few CG water molecules, and therefore positive ions going in one direction are disturbed by negative ions going into the opposite direction. Using AA models, a saturation of the conductivity was also observed above 1~M, compared to experiments, but there is no decrease. A saturation effect is also observed experimentally for conductivity in the bulk for certain types of ions with a large hydration shell~\cite{Bhattacharya2011}.

The CG ionic conductivities values with PW at 320~K are in very good agreement with experimental and all-atom theoretical data for concentrations of KCl below 1.15~M. This result is quite different from ~\cite{Vogele2015} where the authors, using a Nernst-Einstein method, show that conductivity of CG MARTINI ions is close to experiments only if an effective concentration is used.

\subsection*{Charge distribution and electric potential of a DPPC bilayer}

The electric potential across lipid membranes is essential for various biological processes in the cell, such as signaling, ion transport through membrane channels, or translocation of molecules through pores. \emph{In vivo}, such potential difference across the membrane is caused by ionic concentration gradients. 

\begin{figure}[tbh]
\centering
\includegraphics[width=0.7\linewidth]{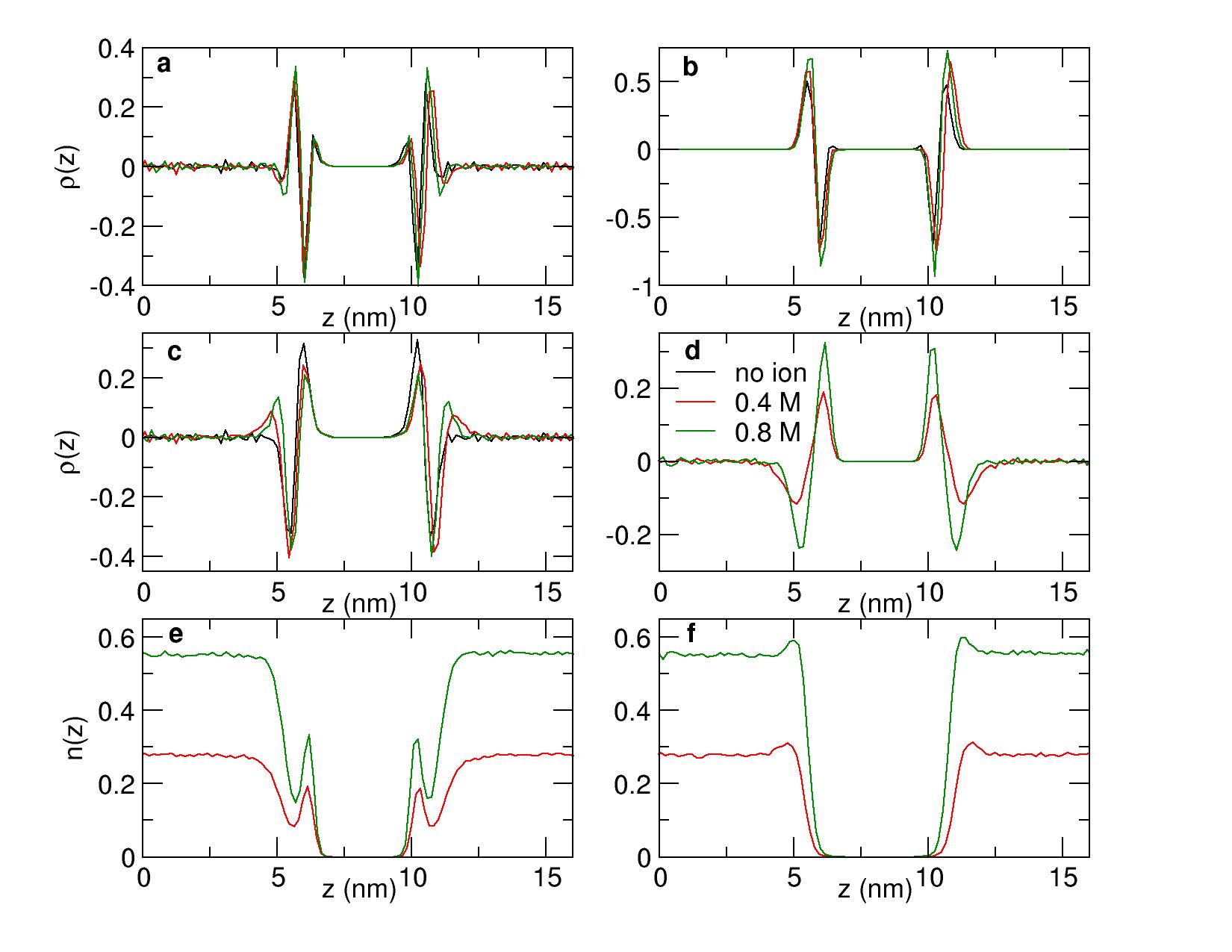}
\caption{(a): Total linear charge density along the $z$ axis around a CG DPPC bilayer with (0.4~M in red and 0.8~M in green) and without ions (in black), and its decomposition: (b): DPPC charge density, (c): solvent, (d): ions. (e-f): Particle density for K$^{+}$ (e) and Cl$^-$ (f) ions.}
\label{fig:densNACL}
\end{figure} 

In nanopore force spectroscopy, this difference is established and monitored using two electrodes placed across the membrane~\cite{Kasianowicz1996}, while there are several options to simulate the trans-membrane potential \emph{in silico}. In this work, a uniform electric field $E$ in the $z$ direction perpendicular to the membrane surface is applied to every charged atom of the system~\cite{Roux2008}. This  method has been previously used by Aksimentiev and Schulten~\cite{Aksimentiev2005} on the $\alpha$-hemolysin pore, and successfully applied to various biological problems~\cite{Pezeshki2009,Modi2012, Bockmann2008, Mathe2005}. An extensive MD study on several membrane systems showed that this method accurately represents an electric potential difference of $\Delta V=EL_z$, where $L_z$ is the system size in the $z$ direction, independently of the membrane thickness~\cite{Gumbart2012}.

In the original paper on the PW model~\cite{Yesylevskyy2010}, the electric potential around a DPPC membrane was computed without any ions and external electric field. Therefore, a study of the electrostatic properties of a system composed of MARTINI \{DPPC+PW+IONS\} under an external electric field is needed to assure the validity of the model for describing the ionic transport. 
After equilibration of a system composed of DPPC lipids, PW water molecules and ions (see Methods), the MD production was performed in the NPT ensemble for two microseconds at 320~K and 1~bar. Several ionic concentrations were simulated: pure water, 0.2~M, 0.4~M and 0.8~M. Moreover, at each concentration, three different electric fields were applied: no external field, 0.01~V/nm  and 0.02~V/nm in the $z$ direction. These electric fields correspond to a theoretical transmembrane potential difference of 0~mV, 160~mV and 320~mV, respectively. It should be noted that, in the NPT ensemble, the box dimension in the $z$ direction increased at the beginning of the dynamics, and the $L_z$ length was around 16~nm on average during all the membrane MD (from 16.24~nm without ions to 16.41~nm at 0.8~M).

\begin{figure}[htb]
\centering
\includegraphics[width=0.8\linewidth]{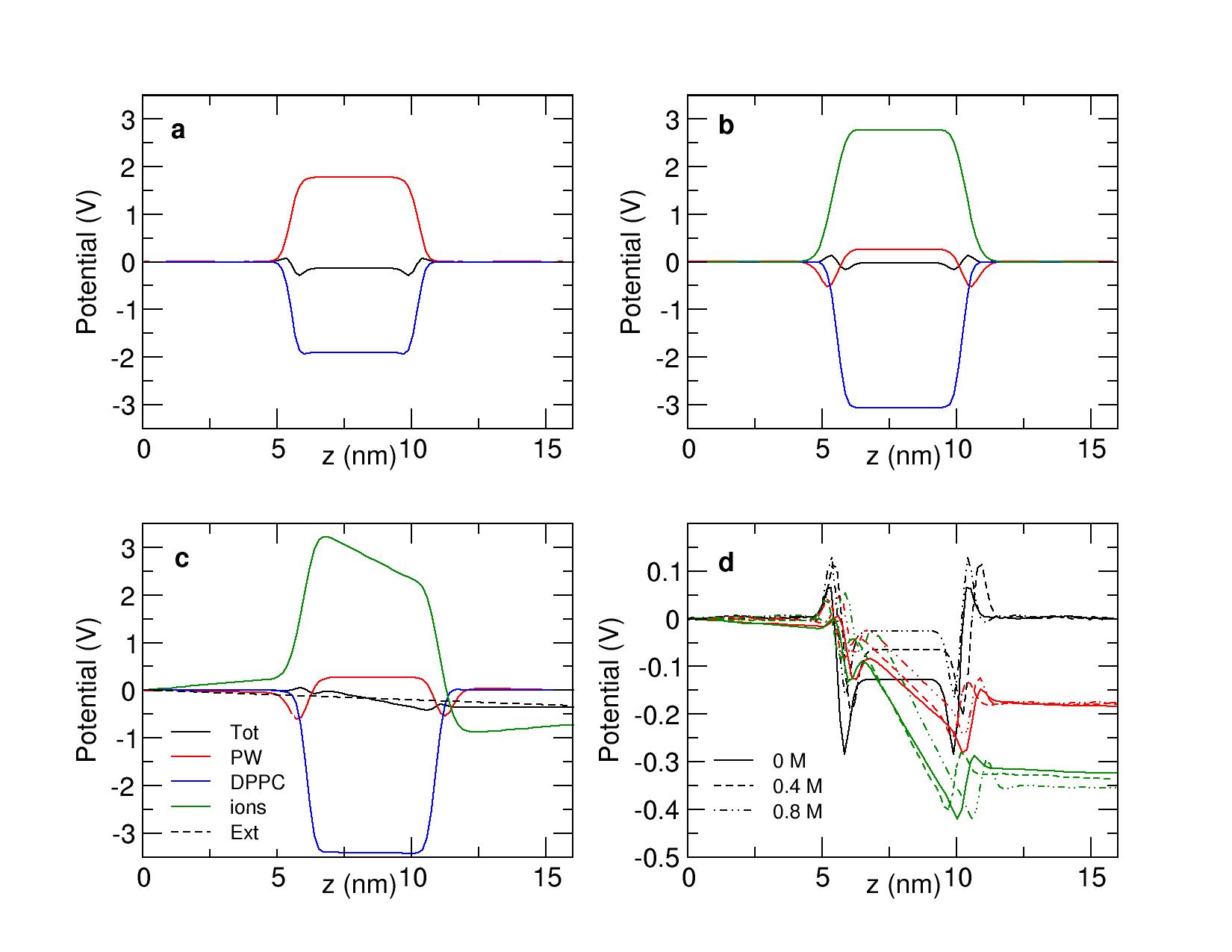}
\caption{(a-c): Total electric potential along the $z$ axis of a CG DPPC membrane (black) and its decomposition, without ions (a), with 0.8~M ionic concentration (b), with no external electric field, and in the presence of an external electric field of 0.02~V/nm (c). (d): Total electrostatic potentials with no electric field (black), electric fields of 0.01~V/nm (red) and 0.02~V/nm (green) without ions (solid line) and with 0.4~M and 0.8~M concentrations (dashed lines).}
\label{fig:potNACL}
\end{figure}
The averaged charge density, on the last 500~ns of the 2~$\mu$s-simulation along the $z$ direction for each species, DPPC, water and ions, was computed for each simulation. Fig.~\ref{fig:densNACL}a shows the total linear charge density and its decomposition (b-d), without an electric field, as a function of the $z$ coordinate, without ions and at 0.4~M and 0.8~M concentrations. The total charge density is globally similar to previous AA simulations~\cite{Rodriguez2011}, showing four peaks from either side of the membrane. The decomposition of the charge density is also consistent with the AA one, so that water and ions, when present, rearrange in reaction of the membrane charge density. The DPPC charge density (Fig.~\ref{fig:densNACL}b) is slightly different when ions are present in the system, showing that the lipids rearrange a little in the presence of salt. The solvent charge density is very different with and without the presence of ions (Fig.~\ref{fig:densNACL}c), since the dipolar moments of water molecules react to the ionic charge distribution. Fig.~\ref{fig:densNACL}e-f show the particle density for each ions species (K$^{+}$ (e) and Cl$^-$ (f)). From these figures, we can infer that there is a layer of cations close to the negative charges of the lipid heads, and an overlay of anions close to the positive beads of the lipid heads. This ionic structure on the membrane surface is also consistent with AA simulations.

The electrostatic potential along the $z$ direction for each was then calculated by the double numerical integration of the charge density, according to~\cite{Bockmann2003,Lee2008}. In the absence of an external electric field, the system, centered on the bilayer, is theoretically symmetrical in the transformation ($x$,$y$,$z$) to ($x$,$y$,$-z$). Therefore, for the systems without electric field, the charge density $\rho(z)$ was symmetrized before calculation of the potential. Fig.~\ref{fig:potNACL}a shows the decomposition of the electric potential without ions. When no ions are present, the PW particles are oriented to compensate the electric potential created by the lipids. However, the total electric potential inside the membrane is negative, contrary to AA~\cite{Rodriguez2011} for which the total electric potential is slightly positive inside the membrane. The PW solvent electric potential does not compensate enough the potential created by the DPPC membrane, as it was previously shown in the original PW paper~\cite{Yesylevskyy2010}.

On Fig.~\ref{fig:potNACL}b, the decomposition of the electric potential at 0.8~M ionic concentration is shown. The total electric potential inside the membrane is less negative than without ions. The potential created by the membrane increases with the ion concentration, as in AA simulations~\cite{Rodriguez2011}, proving that the bilayer is sensitive to the ions presence. When ions are present, the solvent potential is much lower and the rearrangement of ions is compensating the lipids potential. This effect is consistent with the AA study by Rodriguez and Garcia~\cite{Rodriguez2011}, that showed that in the presence of ions, water plays a significantly smaller contribution to the total electric potential. Therefore, although the total electric potential inside the membrane is negative instead of positive, CG ions and water adopt the same behavior around a lipid membrane as an AA force-field.

Fig.~\ref{fig:potNACL}c shows the same decomposition for simulations with an external electric field. The total potential thus also contains the applied external potential (in dashed line). The figure shows that the ionic polarization potential is strongly altered by the external potential. This is not the case for other species whose charge density distribution is much less modified by the external field. On Fig.~\ref{fig:potNACL}d, the total electrostatic potentials of all simulations are represented. In the bulk, the reaction polarization potential is opposed to the external imposed one, and the resulting total electric potential difference is therefore only effective over the membrane. This result is consistent with previous AA-MD~\cite{Gumbart2012} studies which attest that the electric potential does not depend on the thickness of the membrane. The authors showed that the reaction potential of the solution, composed of solvent and ions, to the external field, combined to the external electric field, results in an electric potential difference across the membrane only, with a zero-potential difference within the two bulk regions, proving that using the relation $\Delta V=EL_z$ is accurate. Our results show that this assumption is also accurate with the MARTINI CG model.

We showed that the MARTINI force field with the PW solvent reproduces accurately the ionic conductivity in the bulk and the electrostatics properties around a lipid membrane in the presence of an external electric field. We can therefore now try to simulate ionic transport through a nanopore using this methodology.

\subsection*{Ionic current through the nanopore}

\begin{figure}[htb]
	\centering
    \includegraphics[width=0.4\linewidth]{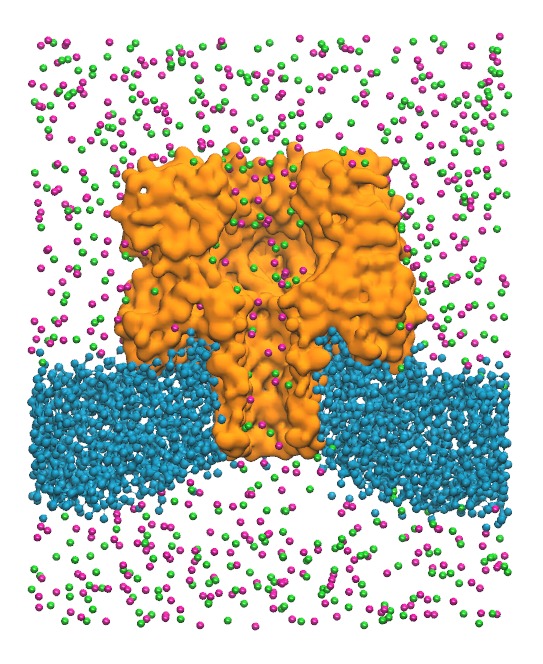}
    \caption{\label{memb_courb} Representation of simulation setup. Water has been removed for clearness purposes.
    Section of the pore (\emph{orange}) along the Z axis. The DPPC membrane (\emph{blue}) is curved. K$^+$ (\emph{green}) and Cl$^-$ (\emph{magenta}) ions are going through the pore channel.}
    \end{figure} 

In this section, we present the results of the simulations for the system composed by the $\alpha$-hemolysin (PDB entry: 7AHL~\cite{Song1996}) inserted in a DPPC bilayer and surrounded by PW water, ions and counter-ions in the presence of an external electric field. A value of 1,1~M$\pm~0.1$~M ionic concentration was measured at the bulk region. Further details of these procedures and the parameters used can be found in Methods.

The initial condition for the ensemble $\alpha$-hemolysin suspended in a lipid bilayer membrane was taken from final state of Prof. Sansom simulations~\cite{Scott2008}, where the lipid bilayer grows spontaneously around the protein. This is the reason for the observed membrane curvature in the pore vicinity in our system (see Figure~\ref{memb_courb}). MD simulations of $\alpha$-hemolysin in DMPC lipid membrane using MARTINI force field were recently studied~\cite{Desikan2017} and showed a considerable distortion of the protein stem. In our simulations, the ElNeDyn network\cite{Periole2009} was applied inside and between the different protomers, and, as a result, the protein stem remains stable during the whole simulation.

After equilibration of the system, production simulations were carried out in the NPT ensemble ($T=320$~K and $P=1$~atm), with an external electric field applied perpendicular to the lipid layer, ranging from $-0.04$~V/nm to $+0.04$~V/nm. The potential bias $\Delta V$ value for a given electric field is calculated using an average $L_z=18.5$~nm, $\Delta V = L_z E_z $. This $L_z$ value varies less than $5~\%$ during the simulation time. All simulations, including the no-field simulation, start from the same initial condition (see Methods). The complete list of simulations performed with their corresponding electric fields and simulation times are shown in Table~\ref{TableSimPore}. 

To provide reasonable statistics, we performed simulations of 2~$\mu$s and 3~$\mu$s, depending on the applied external electric field. We do not present data for an electric field $E_z=-0.04$~V/nm, since the membrane broke early during the simulation. This is also the case in experiments where the allowed potential differences are $|\Delta V| < 350~mV$.

The instant current was calculated computing the number of cations and anions ($n_{c}(t)$ and $n_{a}(t)$) crossing a $xy$ plane perpendicular to the $z$ axis, with $z$ abscissa equal to the position of the center of mass of the groups in the \textit{trans} end of the pore, between $t$ and $t+\Delta t$ ($\Delta t = 100~ps$) (see Supplementary Information). The sign represents the direction of the velocity in the $z$ axis. These values are normally 0 or 1, so we define the number of ions passing through the plane in the interval $[0,t]$ as $ N_{c,a}(t) =\int_0^t n_{c,a}(t') dt' $, and the total charge going through the pore, which is the cumulative function of the current, is : $Q(t)= e \left( N_{c}(t)-N_{a}(t) \right) $.  The instant current is defined then as the slope of the function $Q(t)$ : $I(t) = \frac{dQ(t)}{dt} $. 
\FloatBarrier
\begin{table}[htb]
\centering
\caption{Simulation parameters and number of ion passages during simulations. Electric field  $E_z$, computed potential bias $\Delta V$, total simulation time $t_s$, number of cations and anions traversing during the total simulation time $N_c(t_s)$ , $N_a(t_s)$  and total charge traversing during the simulation time $Q(t_s)$.}

\label{table}
\begin{tabular}{|r|ccccc|ccccc|}

 \hline
\bfseries  &\multicolumn{5}{|c|}{\textbf{Positive Bias}} & \multicolumn{5}{c|}{\textbf{Negative Bias}} \\
\hline
   $E_z$ (V/nm) & $\Delta V (mV)$ & $t_s (\mu s)$ & $N_ {c}(t_s) $ 
& $N_{a}(t_s) $ & $Q_{ts}  (e)$ & 
$\Delta V (mV)$ & $t_s (\mu s)$ & $N_ {c}(t_s) $ 
& $N_{a}(t_s) $ & $Q_{ts}  (e)$

\\
  \hline
$\pm$ 0.00& 0 & 2 & -4 & 6 & -10 &  &  & &  &  \\

$\pm$ 0.005 & 92 & 3 & 45 & -108 & 150 & -92 & 3 & -21 & 103 & -124 \\

$\pm$ 0.01 & 185 & 3 & 85 & -180 & 265 & -185 & 3 & -33 & 209 & -242 \\

$\pm$  0.015 & 277 & 3 & 118 & -212 & 330 & -277 & 3 & -31 & 269 & -300 \\

 $\pm$ 0.02 & 370 & 2 & 144 & -228 &  372& -370 & 2 & -28  & 218 &- 246 \\

 $\pm$ 0.025 & 462 & 2 & 80 & -130 & 210 & -463 & 2 & -26 & 178 & -204 \\

 $\pm$ 0.03 & 555 & 2 & 109 & -113 & 222 &-555 & 2 & -32 & 194 & -226 \\

$\pm$ 0.04 & 740 & 2 & 241 & 229 & 470 & -740 & X & X & X & X \\

\hline

\end{tabular}
\label{TableSimPore}
\end{table}

The number of ions crossing the pore during the whole simulation time, $N_{a}(t_s)$ and $N_c(t_s)$ are represented in Table~\ref{TableSimPore}, together with the total charge variation during the whole simulation $Q(t_s)$ in \emph{e} units. These values are in the limit of "reasonable" statistics although, as we will see below, they evolve with time. The rectification and selectivity of the pore can be also deduced from this global values.

\begin{figure}[htb]
	\centering
   \includegraphics[width=0.9\linewidth]{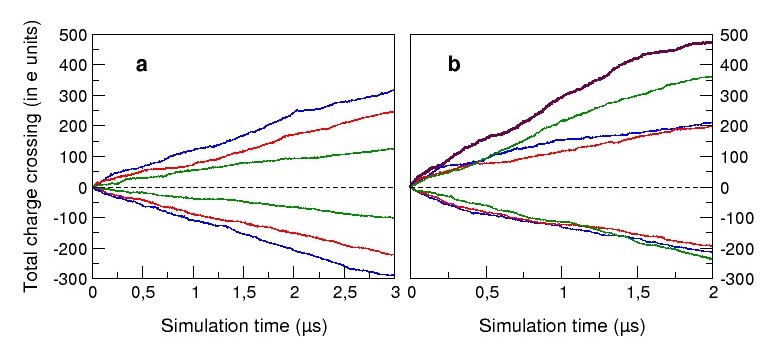}
	\caption{Cumulative current as a function of time, for potentials (a) under or (b) over 300~mV computed during $3~\mu s$ or $2~\mu s$, respectively. Fig. a) : $\pm$ 92~mV (green),  $\pm$ 185~mV (red) and  $\pm$ 277~mV (blue). Fig. b) : $\pm$ 370~mV (green),  $\pm$ 462~mV (red), $\pm$ 555~mV (blue) and  $\pm$ 740~mV (violet). }
    \label{fig:currentCumul}
\end{figure}

The instant cumulative current $Q(t)$ is represented in Fig.~\ref{fig:currentCumul} for potentials under (side a)) and over (side b)) 300~mV. Since the instant current $I(t)$ is the slope of this function at each time $t$, the linear increase of the cumulative current with time would indicate a stationary current. This seems to be the case for potential bias under 300~mV, although it should be pointed out that the stationary current is established only after some hundreds of nanoseconds, especially in the case of positive potential bias. The fact that the slope of the function increases with increasing potential is the reflect of a linear-response regime.

This is not the case for potentials over 300~mV, as shown in Fig.~\ref{fig:currentCumul} b). The cumulative currents are not proportional to the potential bias at the end of the simulation, implying that the simulated systems are not in the linear response regime. Moreover, we cannot be confident that the stationary state has been reached at the end of some simulations as the slope of the cumulative current is still varying. 

\begin{figure}[htb]
	\centering
    \includegraphics[width=0.7\linewidth]{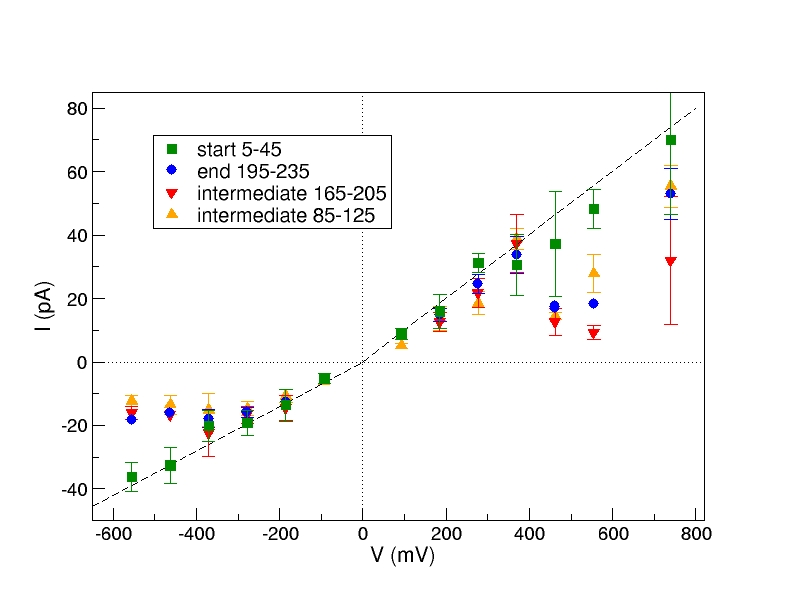}
    \caption{Average current during different passages windows as a function of potential bias. Note the difference between the begining of the simulation (5-45 crossings) and the end of the simulation (195-235 crossings). The black solid line corresponds to $I^{ref}/10$.}
    \label{fig:CurrentIV} 
\end{figure}

Fig.~\ref{fig:CurrentIV} shows the evolution in time of the $I-V$ curve of the system averaged at a fix number of crossings ($N= 40$ events). For every simulation, an initial time $t_{ini}$ is defined as $Q(t_{ini})= 5~e$, which is the time when five charges have already crossed the pore. From this origin $t_{ini}$, we computed the total time $t_{end}$ such as $\Delta Q=Q(t_{end})-Q(t_{ini})= 40~e$. In this interval, the current $\Delta t=t_{end}-t_{ini}$, is estimated as $\bar{ I} = \frac{\Delta Q}{\Delta t}$. The error bars were calculated with the standard deviation $\sigma=\frac1N
\sum_{i=1..N} | \Delta Q_i-I \,\Delta t_i| $ so that 
$\delta I = \frac{2\,\sigma}{\Delta t}$.

In order to compare our results to experimental values, we have defined a reference current in the linear regime, as $
 I^{ref} = R^{ref} \Delta V  $, 
where $R^{ref}= 1~G\Omega$ is a typical resistance observed experimentally for these systems~\cite{Bhattacharya2011}. For negative electric fields, $I^{ref}$ value is multiplied by 0.7 to take into account the observed rectification of $30~\%$ in experimental setups. The black continuous line observed in the figure is the function $I^{ref}/10$. Thus, in our simulations, we observe a $Q_s$ value ten times smaller than the expected one. This is a consequence of the CG model: the resistance of the pore increases since the mobility of the ions inside the channel is highly reduced because of the larger size of PW water molecules. This ratio increases for higher potential differences where the linear regime is not valid anymore. 

The measured current is clearly evolving during our simulations. In spite of this, a linear response regime can be established for potential biases below $\sim 350$~mV.  In this regime, the final current appears slightly smaller than the one measured at the beginning of the simulation and, as mentioned, it is one order of magnitude smaller than the experimental one. For potentials above this value, a limiting current is observed. We find some asymmetry between positive and negative biases: while, for negative biases, the limiting current seems to converge to a well defined value, for the positive ones, this value appears much more noisy. 

The origin of the non-linear ohmic behavior of the current-voltage curve has been explained as due to the crowding or depletion regions at the \textit{trans} and \textit{cis} ends of the pore in presence of an applied external potential~\cite{Tagliazucchi2015}. Both limiting and over-limiting currents in nanopores have been discussed in van Oeffelen \textit{et al}~\cite{vanOeffelen2015}. In our simulations, we observe a limiting current at a quite low voltage $\sim 350$~mV. This value is probably higher in AA simulations, although it is difficult to observe as the simulated time is not enough to get to the steady state~\cite{Aksimentiev2005}. In Ref.~\cite{Liu2013}, the authors compare linear-response theory with AA-MD simulations to determine the departure from the linear regime for two pore models, one of them being a modified $\alpha$-hemolysin pore. Their conclusion, and also ours as regard of our results, is that $\alpha$-hemolysin has a small linear response range and that currents at low voltages cannot be safely extrapolated from the values obtained at higher potential biases. The limiting current cannot be experimentally observed since, as mentioned, the membrane breaks over $\sim 300$~mV .

It is not clear from our results if an over-limiting current appears over $+700$~mV~\cite{vanOeffelen2015}. This could be the case but it cannot be excluded that the stationary state has not been reached, that the effective protein charge (protein plus decorating ions) evolved, or that the periodic boundary conditions are inducing such phenomenon.

As observed, the average current cannot be accurately computed before reaching the stationary state and therefore,  an interesting issue is which simulation time is needed to get to this regime. In van Oeffelen et al.~\cite{vanOeffelen2015}, it is pointed out that, due to slow co-diffusion of ions, the time for the charge-current to decay to a stationary state value is determined by the diffusion of particles. This time is estimated as a typical diffusion time $\tau=L^2/D$, where $L$ is a typical distance and $D = 2 (D_{+}\, D_{-})/(D_{+}+D_{-})$ is an averaged diffusion coefficient for cations and anions. In our simulations, this time is estimated to $\sim 140~ns$, where we have used $L$ the length of the system occupied by the solvent, $D_{+}=2.54~10^-9~m^2/s$ and $D_{-}=2.45~10^-9~m^2/s$ measured during our conductivity calculations. We have computed the evolution of the number of ions inside the stem, the region of the pore going from the trans side to 4~nm above (see Supplementary Information).We observed that the time to get to a stationary number of ions goes from $\sim$ 100 to 300~ns, being shorter for negative voltage biases. The asymmetry for positive and negative biases is in agreement with AA simulations, as well as the number of ions in the stem computed in previous AA and PNP studies~\cite{Noskov2004,Cozmuta2003}.

The time to get to a stationary regime is longer than this estimation according to our simulations. The relatively quick convergence of the number of ions inside the pore is not a guarantee for the stationary state, where anions and cations currents, or conversely charge-current and particle-current, should be, in average, time independent. This is not the case for all our simulations. We observe the convergence of the current for voltage biases under $+350$~mV as well as for negative higher biases. Positive high biases would require longer simulations to get confident results. We can conclude that a 1-1.5$\mu$s simulation should be enough to get a stationary state in the linear regime. 

A final pertinent question about CG-MD simulations of such complex systems is if the pore dynamics play a determining role on the current. If this was not the case, simulation time could be saved by tabulating the pore and membrane potential. To test this possibility, we have performed CG-MD simulations where the backbone of the $\alpha$-hemolysin was constrained to a fixed configuration. The external electric fields imposed were $\pm~0.01$~V/nm, $\pm~0.02$~V/nm and $\pm~0.03$~V/nm. The measured current in this case is much smaller than the corresponding one for the flexible protein and we found clogging periods during the simulation, going up to 200 ns. The current saturates for small potential biases going to a value close to zero (see the Supplementary Information). This result proves that the flexibility of the protein is essential to simulate the translocation process. 

\section*{Discussion}

The use of MARTINI force field to simulate the transport of ions through the $\alpha$-hemolysin in a DPPC bilayer is studied for the first time, to our knowledge, in this article.

In first place, we have investigated the transport properties of the original Martini hydrated ions as a function of concentration. This leads us to the problem of comparing CG results to AA or experimental ones. Since the ratio ion/water valid for AA systems is no longer a suitable parameter to set a molar concentration for the CG simulations, we have decided to use the number of ions per volume as concentration. We are aware that this definition may not represent exactly the same concentration in NVT and NPT ensembles, and definitely cannot be directly compared to AA results. In spite of this incertitude, we can conclude that MARTINI ions in PW water present a conductivity that could represent a KCl solution below 1.1~M concentration. 

Secondly, we have investigated the properties of the DPPC+IONS+PW system in the absence and under different applied electric fields. It is already known that the MARTINI force field leads to a unrealistic positive potential inside the membrane in the absence of an applied electric field~\cite{Yesylevskyy2010}. In the presence of an external field, we found that the expected potential differences through the membrane are recovered, such as in previous AA simulations~\cite{Gumbart2012}. We have also pointed out the role of the polarization of water to describe the correct electrostatics, proving that it is necessary to use a polarizable model of water. 

Finally, we have studied the whole system $\alpha$-hemolysin+DPPC+PW+IONS under different applied external fields focusing on the ionic transport through the pore. Principally due to the size of MARTINI PW water, the pore presents a smaller conductance than the experimental system. This deviation seems to be systematic for low applied external fields. The qualitative behavior of the system nevertheless appears to be correct. Our $I-V$ curves show a small range $\pm~350$~mV where the linear response regime is established. For higher potential biases, a saturation of the current is observed. These limits cannot be compared to experiments, for practical reasons, or to AA-MD simulations, since the simulated time is not enough as to get to a stationary state. We have also performed simulations to conclude that the flexibility of the protein channel plays an important role.

PNP solutions have provided to be extremely useful to describe the stationary state of ionic currents in many systems, but a precise description of this particular system would require excluded volume effects, water-ion correlations, hydrodynamics, finite system size correlations and the possibility to take into account the flexibility of the protein. On the other side, AA-MD simulations provide good current statistics at reasonable simulation times for high voltages but a) reaching the steady state seems to take a long time, if it is possible due to the finite system size, and b) the system is at a non-linear regime. Although the non-linearity will appear at higher potentials in AA simulations than in CG simulations, in Ref.~\cite{Liu2013}, they estimate this limit to be below $500$~mV in a AA-MD study of a modified $\alpha$-hemolysin, interpolating low voltage values from high voltage ones can be somehow risky. At low voltages, direct sampling of net ion fluxes with AA simulations would require very long simulation times to acquire sufficient statistics.  

Our conclusion, as regard to our results, is that CG-MD simulations could result a good alternative to study the details, such as mutation effects, of translocation through protein nanopores, saving some of the simulation time required for the equivalent AA simulations for these kind of problems.

\section*{Methods}

\subsection*{Simulation details}
We performed Molecular Dynamics (MD) using the Gromacs software package~\cite{VanDerSpoel2005} and the MARTINI 2.2 CG model for ions, lipids~\cite{Marrink2007}, proteins~\cite{Monticelli2008,Jong2013} and the PW (Polarizable Water) model~\cite{Yesylevskyy2010} for water. Conductivity simulations systems contained only water and ions at different concentrations, whereas membrane simulations systems contained a DPPC bilayer, water, and ions at different concentrations. In the nanopore simulations, the protein nanopore was inserted in the lipid membrane.\\

A timestep of 20~fs was used for all MD simulations, which is the recommended timestep for CG MD simulations using MARTINI force field. Periodic boundary conditions in the three dimensions were applied. All MD were performed using the PME method for electrostatic interactions, with a 2~\AA\ spacing for the Fourier grid and a direct space cut-off radius of 13~\AA. A relative dielectric constant of $\epsilon_r=2.5$ was used for PW as recommended\cite{Yesylevskyy2010}. Van der Waals interactions were shifted at 13~\AA. A Nos\'e-Hoover thermostat was used to maintain the temperature at 320~K.

\subsubsection*{Conductivity simulations}

For conductivity simulations, the system was originally built as a cubic box of 15.8~nm$^3$, containing 32,756 PW water particles. The system was minimized and slightly equilibrated (2~ns at NVT) before the introduction of ions at different concentrations. The \emph{genion} tool of Gromacs was used to replace random water particles by positive (the MARTINI Na$^+$ ion) and negative (Cl$^-$) ions in the same quantity, in order to keep the system neutral. The system containing water and ions was then again minimized and equilibrated with a short MD (10~ns) in the NVT ensemble, using the Berendsen thermostat. Then, a one-microsecond-long MD was performed in the NVT ensemble at 320~K, using the Nos\'e-Hoover thermostat, with a characteristic time interval $\tau_T=1$~ps for both groups (ions and water). 
For each concentration, ranging from 0.1~M to 2~M, three different simulations were performed: one with no external electric field, one with an electric field of 0.02~V/nm in the $z$ direction, and one with an electric field of 0.04~V/nm in the same direction. 

\subsubsection*{Membrane simulations}
As for the membrane simulations, a box containing 128 DPPC CG lipids organized in a bilayer was downloaded from the MARTINI website~\cite{MARTINIlipids}. 4,232 water molecules (PW) were added around the membrane in a box of originally $6.7 \times 6.1 \times 12.0$~nm$^3$, the $z$ axis oriented perpendicularly to the membrane plan. The system was minimized and shortly equilibrated before addition of ions at different concentrations using the same procedure as the conductivity simulations. Simulations without ions were also performed using the same methodology. After another minimization of the system, a NVT MD simulation was performed for 2~ns using the Berendsen thermostat, followed by 120~ns of NPT MD using the Berendsen thermostat and Berendsen semi-isotropic barostat at 1 bar. The MD production was performed for two microseconds using the Nos\'e-Hoover thermostat at 320~K (with a characteristic time interval $\tau_T=1$~ps for each group) and the Parrinello-Rahman barostat to keep the pressure at 1 bar (with a characteristic time interval of $\tau_P=5$~ps and a compressibility of $4.5\times 10^{-5}$~bar$^{-1}$). It should be noted that, in the NPT ensemble, the box dimension in the $z$ direction increased at the beginning of the dynamics, and the $L_z$ length was around 16~nm on average during all the membrane MD. In addition to simulations without ions, three different KCl concentrations were simulated: 0.2~M (60 ions of each species), 0.4~M (124 ions) and 0.8~M (248 ions). For each concentration, three different simulations were performed: one with no external electric field, one with an electric field of 0.01~V/nm in the $z$ direction, and one with an electric field of 0.02~V/nm. The averaged charge density was calculated on the last 500~ns of the 2~$\mu$s-simulation, using the \emph{g\_potential} tool of Gromacs, decomposing the system into 100 slices in the $z$ direction.

\subsubsection*{Nanopore simulations}
 
The crystallographic structure of $\alpha$-hemolysin was taken from the PDB (entry 7AHL~\cite{Song1996}). The coordinates of the missing atoms were added using the \emph{pdb2pqr} software and the crystallographic water molecules were remove.
The protein was reduced in coarse grains using the \emph{martinize.py} script. The ElNeDyn MARTINI force field was applied on the protein~\cite{Periole2009}. It is composed of an elastic network model that maintains the secondary structures during the dynamics using springs (500~kJ.mol$^{-1}$.nm$^{-2}$ elastic bond strength and a 0.9~nm cutoff).
After a short energy minimization, the resulting CG protein structure was inserted in a DPPC bilayer using the following procedure: the PDB structure of a CG $\alpha$-hemolysin inserted in a DPPC bilayer of 756 lipid molecules obtained by Prof. Sansom \cite{Scott2008} was downloaded from CGDB (coarse grain database, a database of inserted membrane proteins inside DPPC bilayers~\cite{CGDBSamson}. Since the protein CG model used for CGDB is different from the MARTINI one, a least-squared fit of the backbone of our CG protein on the CGDB one was performed, then the initial protein from CGDB was deleted to keep only our MARTINI CG protein inserted in the DPPC bilayer. The size of the simulation box was initially set at 15.85 x 15.85 x 20~nm. The system was solvated with 28,758 standard W MARTINI CG water molecules using the \emph{genbox} Gromacs command. Another minimization was carried out followed by two equilibration steps: 1~ns in the NVT ensemble (using a 10~fs timestep) followed by 5~ns in the NPT ensemble (using a 20~fs timestep).\\

The ionic concentration was set around 1~M, corresponding to 2054 ions of each species (calculated according to the number of initial water molecules and using the ions/PW molecules ratio from conductivity simulations concentrations).
Under normal conditions, the nanopore carries a global charge of $+7e$. Therefore, in order to neutralize the system, 2050 K$^+$ ions and 2057 Cl$^-$ ions were added to the system. 
A short dynamics of 5~ns was then performed at NPT. After transformation of MARTINI W water into polarizable PW water, a short minimization was carried out, followed by two additional equilibration dynamics of 1 ns at NVT and 10~ns at NPT, both using position restraints on the protein backbone (with a 20~fs integration timestep). A last dynamics of 1 ns at NPT without position restraints was performed using the PME method for electrostatic interactions. The resulting system was used as initial state for all the simulations of the pore as well as the dynamics with the pore under position restraints on the protein backbone.

Several electric fields were applied in the $z$ direction of the system, perpendicular to the membrane plane, both positive and negative: no electric field, +/-0.005~V/nm, +/-0.01~V/nm, +/-0.015~V/nm, +/-0.02~V/nm, +/-0.025~V/nm, +/-0.03~V/nm and +/-0.04~V/nm. Molecular dynamics were carried out for 2 to 3~$\mu$s (+/-0.005~V/nm, +/-0.01~V/nm, +/-0.015~V/nm). See Table~\ref{table} for further details.

\bibliography{CGNanopore}

\section*{Acknowledgements}
This work was granted access to the HPC resources of CINES under the allocations 2015 - c2015077139, 2016 - c2016077139 and 2017 - c2017077139 made by GENCI.
Part of this work was funded by CNRS, Défi InFIniTi for the DYNANO project (2016 and 2017).

\section*{Author contributions statement}
N.B, D.D. and R.R. conducted the simulations and analyzed the results. All authors reviewed the manuscript. 

\section*{Additional information}
\subsection*{Competing financial interests}
The authors declare no competing interests.

\end{document}


\flushbottom
\maketitle

\thispagestyle{empty}

\section*{Supplementary Information}
\beginsupplement

In the original MARTINI article, a CG ion is presumed to contain a first hydration shell composed by six water molecules. This number is commonly used to set the ratio between PW molecules and ions for a given molar concentration, although for concentration setting purposes, the 4 to 1 standard MARTINI mapping seems to be a better choice. In our NPT simulations, we have computed this ratio together with the molar concentration in the bulk regions. In our NVT simulations, we will consider that one CG ion represents an all-atom ion with three water molecules.  

\begin{figure}[htb]
	\centering
	\includegraphics[width=0.7\linewidth]{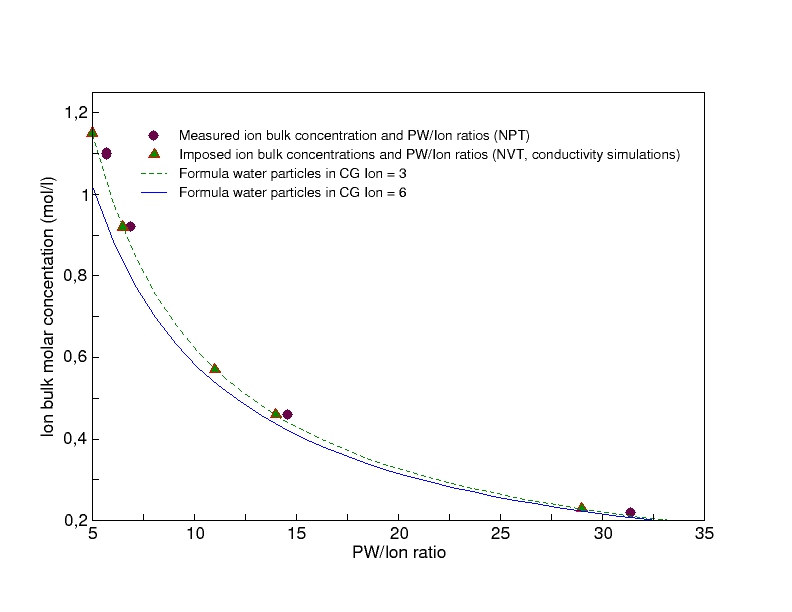}
	\caption{Bulk molar concentration vs. PW/Ion ratio. Circles correspond to measured concentrations in our NPT simulations and triangles correspond to fixed concentrations in our NVT simulations. The dashed line is the theoretical value of molar concentration assuming one CG ion represents an ion plus three AA water molecules. The solid line is the same theoretical value, when a CG ion represents a ion plus six AA water molecules. }
    \label{comp-dens}

\end{figure}  

The instant current was calculated computing the number of cations and anions crossing a $xy$ plane perpendicular to the $z$ axis, with $z$ abscissa equal to the position of the center of mass of the groups in the \textit{trans} end of the pore, colored in blue in the figure. 

\begin{figure}[htb]
	\centering
	\includegraphics[width=0.3\linewidth]{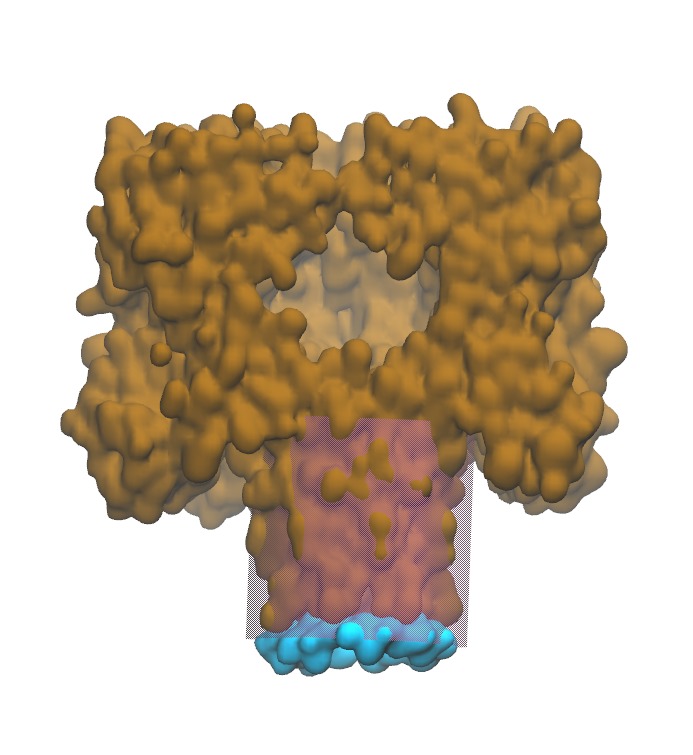}
	\caption{Partial section of $\alpha$-hemolysin. The group in the \emph{trans} side of the pore (\emph{cyan}), composed of Asp127, Asp128, Thr129, Gly130 and Lys131, is used to define the plan position for the current calculations. The evolution of the number of ions was computed inside the restricted cylinder of 4~nm in height and width (\emph{mauve}) surrounding the stem.}
    \label{bott_group}
\end{figure}  

\begin{figure}[htb]
	\centering
    \includegraphics[width=0.5\linewidth]{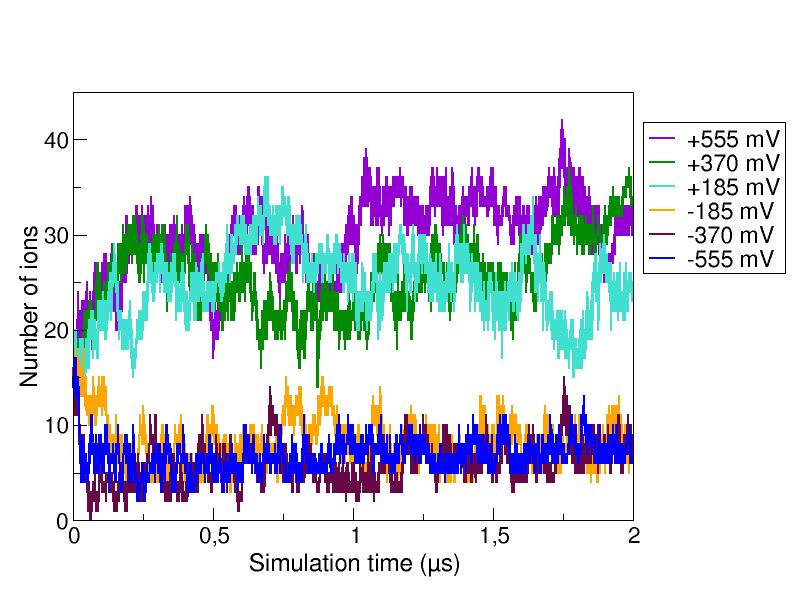}
    \caption{Number of ions in the stem region of the pore along time for different external electric fields applied on the system.}
    \label{fig:countinpore} 
\end{figure}  

We have defined the "stem" as a region inside the pore going from the $z$ abscissa defined as the "trans" to 4~nm above. It is colored in mauve in the figure. 

The evolution of the number of ions inside the stem for different positive and negative biases is represented in \ref{fig:countinpore}. The asymmetry between positive and negative bias can be observed. We have estimated to 1~$\mu$s the time to get to a stationary state.

During the simulations where the protein was constrained to a fix position, the number of ions crossing the membrane is not enough as to compute statistics in the way used for the flexible protein.
In order to compute the current, a different approach was used. 

A simple model to describe the current evolution from the initial condition to the stationary state is

\begin{equation}
I(t)=I_{\infty}+\left( I_0 - I_{\infty} \right) e^{- t/\tau}
\end{equation}

So that the total cumulative current is
\begin{equation}
Q(t) = I_{\infty} t -\tau \left( I_0 - I_{\infty} \right)\left(  e^{- t/\tau}
- 1 \right )
\label{fitequation}
\end{equation}

\begin{figure}[htb]
	\centering
    \includegraphics[width=0.6\linewidth]{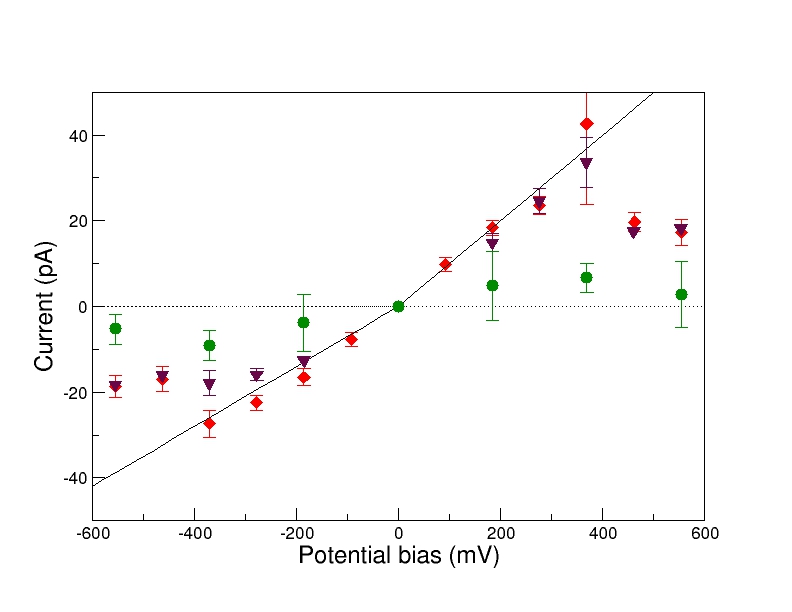}
    \caption{IV curve for flexible protein and rigid fixed protein. Position restrained protein and time statistics (circles), flexible protein and time statistics (diamonds) and flexible protein with crossing statistics (triangles). The black solid line is the reference experimental current, as defined in the main text, divided by a factor of 10.}
    \label{fig:IV-CH-PR} 
\end{figure}

We have fitted the parameters of eq.[\ref{fitequation}] with the computed values of $Q(t)$ for times each $\Delta t = 15$~ns. While $I_0$ and $\tau$ depend on the initial condition, it can be expected, close enough to the steady state, that $I_{\infty}$ is the asymptotic value of the current. To test the model, $I_{\infty}$ has been computed for the flexible protein and compared to the averaged current between $190$ and $230$ crossings. As shown in figure \ref{fig:IV-CH-PR} the results are quite similar between these two methods. It can be observed that the current saturates very quickly going to a value close to zero. We believe that since the flexible-non-fixed protein vibrates and travels along the membrane, it transmits energy to the particles inside the pore and prevents clogging. This is not the case for a rigid fixed protein. The dynamics of the protein seems then essential to simulate the translocation process. 

